\documentclass[traditabstract]{aa}

\usepackage{multicol, blindtext}
\usepackage{graphicx}
\usepackage{txfonts}
\usepackage{natbib}
\usepackage{amssymb}
\usepackage{amsfonts}
\usepackage{amsbsy}
\usepackage{amsmath}
\usepackage{verbatim}

\usepackage{amstext}
\usepackage{pdflscape}
\usepackage{color}
\usepackage{url}
\usepackage{lscape}
\usepackage{longtable}
\usepackage{multirow}

\newcommand{\Kb}{\rm K-band}
\def\HER~{{\it Herschel} }
\newcommand{\be}{\begin{equation}}
\newcommand{\ee}{\end{equation}}
\newcommand{\etal}{et al..\ }

\newcommand{\HII}{\rm H$_2$}
\newcommand{\MHc}{\rm M(H$_2$)c}
\newcommand{\MHv}{\rm M(H$_2$)v}
\newcommand{\XCO}{\rm X$_{\rm CO}$}

\begin{document}

\title{The Molecular Mass function of the Local Universe}
\subtitle{  }

   \author{P. Andreani \inst{1} 	
	\and Y. Miyamoto\inst{2,3} 
        \and H. Kaneko\inst{3,4} 
        \and A.~Boselli\inst{5}
        \and K.~Tatematsu\inst{3,6}
        \and K.~Sorai\inst{7}
        \and R.~Vio\inst{8}
	}

	\institute{European Southern Observatory, Karl-Schwarzschild-Stra\ss e 2, 85748 Garching, Germany, \email{pandrean@eso.org}   
 \and
  National Astronomical Observatory of Japan,  National Institutes of Natural Sciences, 2-21-1 Osawa, Mitaka, Tokyo 181-8588, Japan
  \and 
        Nobeyama Radio Observatory, National Astronomical Observatory of Japan, National Institutes of Natural Sciences, 462-2 Nobeyama, Minamimaki, Minamisaku, Nagano 384-1305, Japan
  \and
        Joetsu University of Education, Yamayashiki-machi, Joetsu, Niigata 943-8512, Japan
  \and
        Aix Marseille Univ, CNRS, CNES, LAM, Marseille, France 
  \and 
        Department of Astronomical Science, SOKENDAI (The Graduate University for Advanced Studies), 2-21-1 Osawa, Mitaka, Tokyo 181-8588, Japan
  \and
    Department of Physics, Faculty of Science, Hokkaido University, Kita 10, Nishi 8, Kita-Ku, Sapporo, Hokkaido 060-0810, Japan
  \and
    Chip Computers Consulting s.r.l., Viale Don L.~Sturzo 82,
              S.Liberale di Marcon, 30020 Venice, Italy
        }

\titlerunning{Local Molecular Mass Function}
\authorrunning{Andreani, Miyamoto \etal}

\date{Received .............; accepted ................}

\abstract{}{We construct the molecular mass function using the bivariate \Kb-Mass Function (BMF) of the {\textit{Herschel}}~\thanks{\textit{Herschel is an ESA space observatory with science instruments provided by European-led Principal Investigator consortia and with important participation from NASA.}} Reference Survey (HRS), a volume-limited sample already widely studied at the entire electromagnetic spectrum.}
{The molecular mass function is derived from the \Kb~ and the gas mass cumulative distribution using a copula method described in detail in our previous papers \citep{PaperI,PaperII}. 
}
{The \HII ~mass is relatively strong correlated with the \Kb~ luminosity because of the tight relation between the stellar mass and the molecular gas mass within the sample with a scatter likely due to those galaxies which have lost their molecular content because of environmental effects or because of a larger gas consumption due to past star formation processes.
The derived \HII MF ~samples the molecular mass range from $\sim 4\times$10$^6$M$_\odot$ to $\sim$10$^{10}$M$_\odot$, and when compared with theoretical models, it agrees well with the theoretical predictions at the lower end of the mass values, while at masses larger than $10^{10}M_\odot$ the HRS sample may miss galaxies with large content of molecular hydrogen and the outcomes are not conclusive.
The value of the local density of the molecular gas mass inferred from our analysis is $\sim 1.5 \times 10^{7}M_\odot $Mpc$^{-3}$, and it is compared with the results at larger redshifts, confirming the lack of strong evolution of the molecular mass density between $z$=0 and $z$=4.}
{This is the first Molecular Mass Function derived on a complete sample in the local Universe, which can be used as a reliable calibration at redshift $z$=0 for models aiming at predicting the evolution of the molecular mass density.}{}

\maketitle

\section{Introduction}\label{intro}
Understanding galaxy formation and evolution is still at its infancy and 
the complete picture that fully describes how galaxies form and evolve into the structures is not yet fully outlined.
One way to address this issue is the careful comparison between the predictions of theories (and simulations) with the observations. The classical approach is the use of the statistical nature of the galaxy properties and determine it over various sets of galaxy populations (i.e. at different redshifts, types, environment, etc).

In many branches of Astronomy and Cosmology a very common tool used to investigate the statistical properties of a population of objects is the luminosity and the mass function \citep[e.g.][]{joh}. In the study of the physical processes that govern galaxy formation and evolution, the stellar and gas mass functions (MFs) in particular constitute one of the fundamental pillars to constrain theories from observations.

Observationally galaxy evolution can be followed studying the baryons (stars and gas) which are supposed to be associated with the dark matter (DM) haloes.
The gas cools and condenses within the haloes, until it produces an independent self-gravitating unit which can form stars, heating and enriching the rest of the gas, and perhaps even ejecting it from the halo \citep{mo}.

In this framework, both hydrodynamical simulations, semianalytical and empirical approaches add the baryonic component to the DM in
haloes, follow their evolution through a merger history including gas and star formation processes.
These star formation and feedback processes are not understood in detail \citep[see for reviews,][]{mo+10,soda}.

To tune these multi-parameter models their outcomes are compared to the statistical properties of galaxies in the local Universe. As mentioned above, key constrains are the mass and luminosity functions (LFs) \citep[i.e.][]{dave,lag+16}.

From the observational side, to compute these statistics, it is necessary to build large samples of galaxies selected at various wavelengths.
The derived LFs and MFs contain some significant uncertainties mainly due to the lack of either the imaging of large fields, or the required multi-wavelength homogeneous coverage and complete redshift information.

\subsection*{Molecular gas and star formation rate}

Observations widely point out that star formation (SF) occurs within dusty molecular clouds \citep[i.e. for reviews][]{sovan,keeva} and that the surface density of the star formation rate (SFR) correlates with the surface density of molecular hydrogen, \HII, roughly following a linear relation, $\sigma({\rm SFR})\propto \sigma$(\HII) \citep[e.g.][]{big,schru}.

The \HII~ content of galaxies has been most studied through the emission from $^{12}$CO as a tracer.
Measuring \HII~ mass requires the definition of the CO-\HII~ conversion factor, \XCO, to convert from its observable tracer, CO, which is itself a function of metallicity \citep[i.e.][]{bol} and stellar mass \citep[i.e.][]{des}.

\subsection*{Why a local molecular mass function?}

Previous molecular MFs of nearby galaxies have been indirectly derived from the CO luminosity distribution built also on assumptions of its relation with other tracers. \citet{ker+03} used an incomplete CO sample based on a far-IR selection
and used the correlation of the CO with the 60$\mu$m luminosity. The resulting CO MF is, therefore, biased towards gas rich galaxies. An updated estimate of
the \HII~mass function (hereafter \HII~MF), based on an empirical and variable CO-\HII~conversion factor, \XCO, was presented by \citet{obra}.
\citet{san+17} and \citep{fle+20} have derived a ${\rm L^\prime(CO)}$ luminosity distribution from the COLD GASS (CO legacy data base for the GASS survey, \citet{san+11}).
This survey, although biased towards massive galaxies (stellar mass, ${\rm M_\star} >$~a~few~$10^{9} M_\odot$, \citet{san+11,san+17}), i.e. it might not sample a sufficiently large dynamic range in ${\rm M_\star} $ to trace a fair distribution, is at present the only survey with a large enough database to allow a fair reconstruction of the ${\rm L^\prime(CO)}$ luminosity distribution. However, this sample too is not unbiased, because it is not blindly selected and its selection from the SSDS survey is prone to many uncertainties due to unknown/uncontrolled systematics.

The \HER~Reference Survey (HRS) is a \HER~guaranteed time key project, performing photometric observations with the SPIRE cameras towards HRS galaxies \citep{bos+10}.
The survey selection criteria (magnitude- and volume-limited, see \S ~\ref{sample}), size and multiwavelength coverage (from UV to radio wavelengths both in spectroscopy and photometry) together with the \HER~ results in the far-IR, sensitive to dust mass down to $10^4 M_\odot$,
have shown that the HRS can be considered as a 'reference' sample to carry out statistical analysis in the local Universe \citep{bos+10,PaperI,PaperII}.

In this letter we compute and discuss the local molecular mass function derived from the HRS sample. CO observations are taken with single dishes and described in detail in other papers \citep{bos+14a,bos+14b,PaperIII}. Our aim is to infer the local molecular mass density on a more solid ground which could be used as a reference to set limits to current models predicting the evolution of the molecular mass with redshift.

The paper is organised as follows. The sample with the additional observations carried out with the FOREST array at the Nobeyama 45m antenna is briefly described in section~\ref{sample}. The computation of the molecular mass function in section~\ref{results} and discussed in section~\ref{discussion}. 

\section{The data}
\label{sample}

The HRS is a volume-limited sample (i.e., 15$<D<$25 {\rm Mpc}) including late-type galaxies (LTGs) (Sa and later) with 2MASS \Kb~ magnitude $\leq$12 {\rm mag} and early-type galaxies (ETGs) (S0a and earlier) with $\leq$ 8.7 {\rm mag}. Additional selection criteria are high Galactic latitude ($b >+55^\circ$)
and low Galactic extinction (AB$<$0.2 {\rm mag}, \citep{sch+98}).
The sample includes 322 galaxies (260 LTGs and 62 ETGs), and the total volume over an area of 3649 sq.deg. is 4539 Mpc$^3$.
The selection criteria are fully described in \citet{bos+10}.\hfill\break
This data set has been extensively used to investigate many aspects of the physics of the interstellar medium, their interaction with the environment, and among others the stellar, dust, gas mass functions, infrared luminosity function and the star formation function  \citep{cor+12a,bos+10,cies+14,bos+14a,bos+14b,bos+15,PaperI,PaperII}.

We have recently investigated the HRS LFs and MFs using the bivariate functions with respect to the K-band luminosity, which is the band at which the sample is complete \citep{PaperI,PaperII}. We have employed an accurate statistical method which makes use of the whole data sets including the upper limits.
We have shown that the HRS samples galaxies down to a few $10^8 {\rm M_\star} $ and thus has a dynamic range sufficiently wide for an accurate determination of the H$_2$ mass function.

We were able to construct the \HII~MF of the HRS sample from the bivariate K-band Luminosity-\HII~mass function because of the good correlation (correlation coefficient, $\rho=0.67$) between the \Kb~ luminosity and the \HII~ mass. Based on this finding we could reconstruct from the bivariate function the analytical form of the \HII~mass function \citep[see for detail][]{PaperI,PaperII}.
Despite all the caveats described in the paper \citep{PaperII}, the molecular mass function is the first function built on a complete sample, although the completeness is in the K-band.

In this work we improve our analysis of \citet{PaperII} making use of additional observations of the CO(1-0) fluxes towards those galaxies previously lacking this measurement. Of the 323 galaxies 225 observations are reported in \citet{bos+14a,bos+14b}, 43 more have been targeted in January-February 2018 and 2019 with the 45 m radio telescope of the Nobeyama Radio Observatory (NRO)\footnote{Nobeyama Radio Observatory
is a branch of the National Astronomical Observatory of Japan,
National Institutes of Natural Sciences.}. Observations will be fully described, analysed and discussed in a paper in preparation \citep{PaperIII}. 

Here we briefly outlined how the observations were taken.
We used the FOur-beam REceiver System (FOREST: \citet{min+16}) and employed the OTF mapping mode. The observed area varied according to the optical size of the galaxies from
$2^{\prime\prime}\times2^{\prime\prime}$ to $4^{\prime\prime}\times4^{\prime\prime}$. The total time for the observations varied between a couple of hours to 8 hours depending on the expected CO line flux.

We used the $2\times2$ focal-plane dual-polarization sideband-separating SIS mixer receiver for the single side band (SSB) operation of FOREST, which provides eight intermediate frequency (IF) paths (i.e., four beams $\times$ two polarizations) independently.
The backend was an FX-type correlator system, SAM45, which consists of 16 arrays with 4096 spectral channels each.
The frequency coverage and the resolution for each array was set to 2 GHz and 488.24 kHz, respectively, which gives a velocity coverage and resolution of $\sim$10000 and 1.5 $km s^{-1}$ at 115 GHz.

At 115 GHz the half-power beam widths of the 45 m telescope with FOREST were $\sim 14^{\prime\prime}$ and it degrades to $\sim 17^{\prime\prime}$ because of scanning effect. The system noise temperatures were 300–500 K during the observing runs. 

 In order to check the absolute pointing accuracy, every hour we observed a pointing source, 3c273, rtvir, r-uma, rleo, using a 43 GHz band receiver. 

The line intensity was calibrated by the chopper wheel method, yielding an antenna temperature, $\rm T^\star_A$, corrected both atmospheric and antenna ohmic losses. The main
beam brightness temperature, $\rm T_{mb}$ was converted from $\rm T^\star_A$ for each IF, by observing a standard source, the carbon star IRC+10216. 
The scaling factors not only correct the main-beam efficiency ($\rm \eta_{mb}$ ) of the 45 m antenna but also compensate for the decrease in line intensity due to the incompleteness of the image rejection for the SSB receiver \citep[e.g.][]{nak+13} (\citep[see][]{sor+19}, for more details).

The new observations refer to those objects with the lowest value of the K-band luminosity and therefore allow us to sample the lowest value of the molecular and stellar masses, down to values of the mass function of $\sim 4\times 10^6$M$_\odot$.
We make use of the total 268 (225 from \citet{bos+14a} and 43 from \citet{PaperIII}) galaxies of which we have 183 detections and 85 upper limits to obtain and discuss the CO distribution function and the molecular mass function derived from the CO(1-0) observations. 

\section{The CO(1-0) luminosity and the molecular mass}\label{results}

The construction of a mass function is very sensitive to the quality of the sample selection. In order to have as much as possible homogeneity across different observations we have recomputed the CO(1-0) line luminosities for all galaxies following \citet{bos+14b}. 

 In the HRS sample we expect that, because of the wide dynamic range in parameters \citep{bos+12}, the conversion factor, \XCO, changes significantly from massive, metal-rich quiescent to dwarf, metal poor galaxies and a constant \XCO~value may underestimate the molecular content at stellar masses below $10^{10}$M$_\odot$. 
We follow then \citet{bos+14b} and compute the molecular mass and the molecular mass function twice, applying a constant and a luminosity dependent \XCO~conversion factor.

The constant conversion factor is \XCO~= 2.3$\times 10^{20} cm^{-2}$/(K km s$^{-1}$) ($\alpha_{\rm CO}$ = 3.6 M$_\odot$/(K km s$^{-1}$ pc$^2$)). The luminosity-dependent \XCO : $\log$ \XCO = -0.38 $\times \log {\rm L_H} + 24.23 [cm^{-2}/(K km s^{-1})]$, where ${\rm L_H}$ are the {\rm H}-band luminosities available from 2MASS for all the HRS galaxies \citep{sk+06} \citep[see for further detail][]{bos+14b}. We refer hereafter to those values as \MHv~ and \MHc.

Following a procedure similar to that applied in previous papers \citep{PaperI,PaperII} we compute the bivariate function of the two values of the \HII~ mass and the
\Kb~luminosity using a copula method \citep[see][]{hof+18}. The difference here with respect to the previous computation is that instead of the gaussianization of the data and the use of the Gaussian copula, the best copula among a set of copula families wasselected by means of a maximum-likelihood approach \citep{hof+18}.
In the present case, the best result is given by a rotated Tawn type 2 copula (180 degrees).

\begin{figure}[h]
\centering
\includegraphics[width=1.1\linewidth]{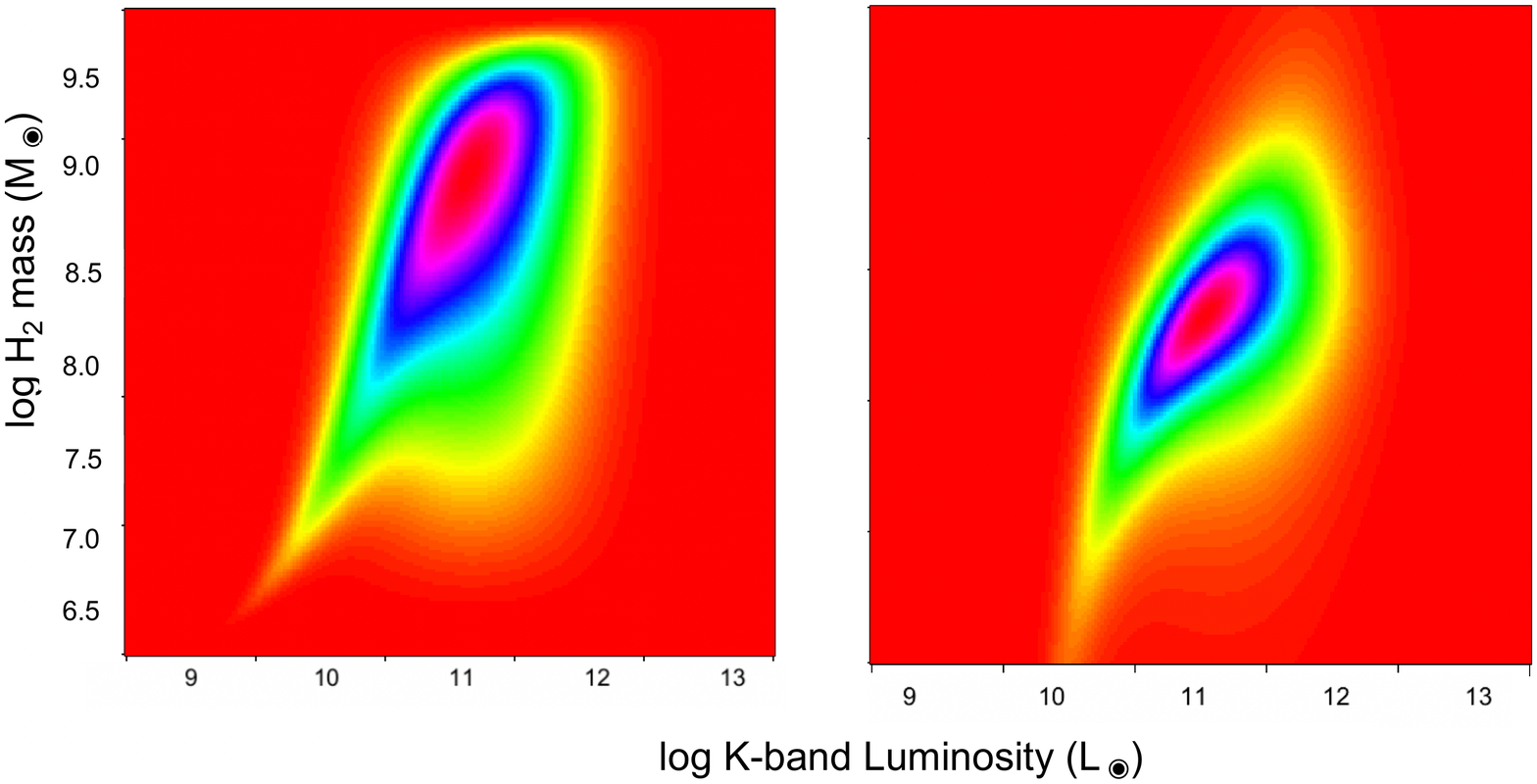}
\caption{The bivariate functions of the molecular masses and the \Kb~luminosity, with the molecular mass derived with a luminosity-dependent conversion factor (left)
and a constant (right) respectively. The colour code refers to the correlation between the two variables (i.e. larger for those lying in the pink region). The spread is larger for the luminosity-dependent conversion factor. In this latter case galaxies with larger \Kb~luminosity have a lower molecular mass content, while those of lower \Kb~luminosity have a large molecular mass content (see text).}
\label{fig:bivariate}
\end{figure}

The bivariate functions of the molecular masses and the \Kb~luminosity are shown in Figure~\ref{fig:bivariate}, for both cases, molecular mass derived with a luminosity-dependent and with a constant conversion factor, respectively. Both functions show a well defined correlation between these two values, which reflects the strong relation between the stellar mass and the molecular gas mass of the HRS sample \citep{bos+14b,PaperII}.

The overall spread seen in Figure~\ref{fig:bivariate} may be due to a slightly lower molecular mass content in HI-deficient galaxies 
\citep{bos+14c,PaperII}. However, the dynamic range and the overall spread are larger for the bivariate function derived with a luminosity-dependent conversion factor. In this latter case we assign a lower molecular  content to those galaxies with a larger \Kb~ luminosity and a larger molecular gas content to those with a lower \Kb~luminosity. 
The ratio between the molecular gas to the stellar mass, M(\HII)/M$_\star$, decreases when the stellar mass increases as seen in the scaling relation in \citet{bos+14b}. This means that in the most massive galaxies the gas was already transformed into stars while in the dwarf galaxies this process has not yet stopped and therefore the gas content is larger at lower stellar mass.

\begin{figure}
{\includegraphics[width=1.1\linewidth]{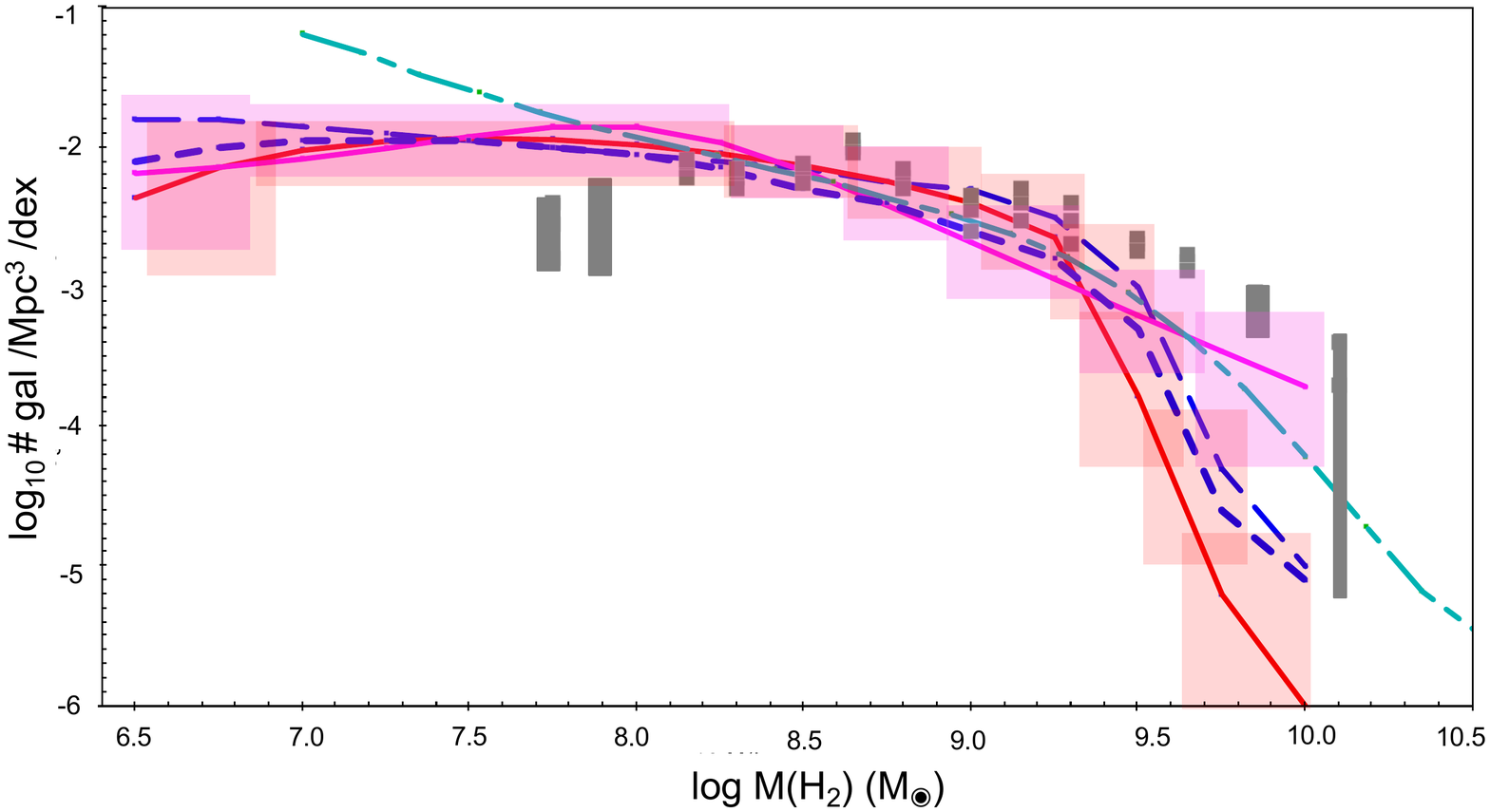}}
        \caption{The molecular mass functions derived in this work using the parameters of the Bivariates \Kb~luminosity-\HII~mass. The red refers to the \HII~MF derived for a luminosity dependent \XCO~ factor, while the pink curve to that derived for a constant \XCO.The blue curves refer to the models reported in \citet{lag+15}, while the green curve to that by \cite{pop+14}. The dark grey bars refer to the values of the \HII~MF by \citet{fle+20}, taken from their fiducial model.}
       \label{fig:MF}
\end{figure}

Figure~\ref{fig:MF} displays the \HII~MF derived from the BMF for the two cases (constant and luminosity-dependent \XCO~in pink and red respectively). The shown curves correspond to the Johnson families used to infer the bivariate function \citep[see detail in][]{vio+94,PaperII}. As widely explained in our previous papers \citep{PaperII} these derived functions are the closest approximation to the analytical form of the corresponding mass functions. We could infer the values of these latter functions down to the lower end of the MFs, i.e. to molecular masses of $\sim 4\times10^6$M$_\odot$.
The errorbars shown in Figure~\ref{fig:MF} were derived by extracting the \HII~MF from the BMF computation allowing the variation of the input variables within their observation errorbars. 
Table~\ref{MFs} reports the average values in every M(\HII) mass bin of the \HII~MF values of the functions extracted from the bivariate computation for the two \XCO~cases (constant and luminosity-dependent \XCO).

We compare the molecular mass functions of the HRS sample with that obtained by \citet{fle+20} for the COLD GASS sample.
This comparison must be taken with caution and it is only indicative. In addition to the issues discussed in the Introduction we lack the information about the galaxy properties to apply the luminosity dependent conversion factor between ${\rm L^\prime(CO)}$ and \HII~equal to the one used in this work \citep{bos+14b} and can be therefore strictly compared only to the curve obtained with the \MHc~values (pink curve in Figure~\ref{fig:MF}).

The new CO(1-0) measurements allow to sample the molecular mass function at a factor of ten deeper than that of COLD GASS.
The discrepancy shown at large mass values can be attributed, on the one hand, to the \XCO~factor. But even in the case of constant \XCO~ the HRS \HII~MF is lower at large masses than the \citet{fle+20}'s values.
The HRS sample misses galaxies with large molecular masses because its volume is too small to contain a significant number of large mass galaxies \citep{bos+14b}. Indeed, the number of galaxies with M(\HII)$\sim 10^{10}~{\rm M}_\odot$ is only 5.
The errorbars at M(\HII)$> 3\times 10^{9}~{\rm M}_\odot$ are corresponding large. Large errorbars are also assigned to the higher end of the MF by \citet{fle+20} and a fair comparison is difficult to make. At low masses the \citet{fle+20}'s curve shows the incompleteness already at M(\HII)$<10^8$M$_\odot$.

In Figure~\ref{fig:MF} we add the predictions of theoretical models, either hydro-dynamical simulations \citep{lag+15} and semi-analytical \citep{pop+14}. It is clear that the HRS sample results put stringent constrains on the available models at the low mass end. While both models by \citet{lag+15} agree with the HRS measurements, that by \citet{pop+14} seems to predict many more low mass galaxies than seen in the local Universe.
To note also is the discrepancy between the predictions of \citet{pop+14} and \citet{lag+15}'s models at M(\HII)$> 6\times 10^9$~M$_\odot$.

\begin{table*}
\caption{The \HII~MF resulting from the bivariate and for the two chosen values of the \XCO~conversion factor}          
\label{MFs}      
\centering
\hskip -0.7cm          
\begin{tabular}{lc c c}     
\hline\hline       
\\
& {M({\rm H$_2$})} &  {H$_2$~MF$_c$} &  {H$_2$~MF$_v$}  \\ 
& {($\log$ M$_\odot$)} &  {(\# gal/Mpc$^{-3}$/dex)} & {(\# gal/Mpc$^{-3}$/dex)} \\

 &  6.5	& -2.19  & -2.36 \\
& 6.75    & -2.14  & -2.14 \\
& 7.0      & -2.08  & -2.02 \\
& 7.25	& -2.01  & -1.96 \\
& -7.5	& -1.92  & -1.93 \\
& 7.75	& -1.85  &  -1.94 \\
& 8.0	& -1.85  &  -1.98 \\
& 8.25    & -1.96  & -2.04 \\
& 8.5	& -2.17  & -2.12 \\
& 8.75	& -2.42  & -2.24 \\
& 9.0	& -2.68  &  -2.40 \\
& 9.25	& -2.94  &  -2.64 \\
& 9.5	& -3.20  &  -3.78 \\
& 9.75	& -3.46  &  -5.2 \\
& 10.0    & -3.62  &  -6.00 \\
\\
\hline           
\end{tabular}
\end{table*}








\section{Discussion}\label{discussion}

The molecular mass function reported in Figure~\ref{fig:MF} is the first function built on a complete sample, exploiting a bivariate analysis of the molecular gas mass and the
\Kb~luminosity. Since this latter mainly traces the star mass, the observed relation is likely dominated by galaxies in which, where stars are located, there is still molecular hydrogen. The scatter seen in Figure~\ref{fig:bivariate} is very likely to be attributed to those galaxies where the molecular hydrogen is deficient as observed in objects strongly influenced by environmental effects and/or in massive galaxies where the star formation activity is low \citep{bos+14c,PaperII}.

The bivariate analysis allows us to derive a \HII~MF of the sample. The MF is compared with the predictions of semi-analytical models \citep{lag+15,pop+14} and shows a good agreement with \citet{lag+15}'s models and a large discrepancy with \citet{pop+14}'s one. The disagreement at large molecular masses 
is only indicative because of the lack of statistical significance in the number of objects in the HRS sample in this mass range. 

It is interesting to understand which insights can be derived by comparing the results of this work with the theoretical predictions. \citet{lag+15}'s models are based on hydro-dynamical simulations using two different prescriptions (GK11 and K13, respectively \citep[see][]{lag+15}) to convert neutral to molecular Hydrogen.
These prescriptions are function of the metallicity (and therefore of the dust-to-gas ratio)
and of the radiation field assumed for the cold neutral medium.

In the GK11 case not all the resulting \HII~gas is locked up in star-forming regions and 20-40 percent may be present in form of diffuse gas not associated with star-forming regions, a fraction of which, however, does not depend strongly on the stellar mass. In the K13 case the fraction of molecular gas locked up in star-forming region is lower (up to 5-20 percent) and depends on the stellar mass for low values of this latter.
The theoretical predictions resulting from the two recipes are both compatible with the observations and the different physical processes implied by the models may be at work in the ISM of the galaxies.
Both models are shown in Figure~\ref{fig:MF}.\hfill\break
\citet{pop+14} adopt a semi-analytical model with two different approaches for calculating the molecular fraction of the cold neutral gas in a galaxy. The first is based on an empirical recipe dependent on the gas pressure, while the second is the same GK11 used by \citet{lag+15}. The number of galaxies at low and large \HII~gas mass lies far above the observed number densities.
While the observational uncertainties at large masses are quite significant and larger complete samples containing data of massive galaxies need to be built before any conclusion can be drawn, the disagreement at low masses is large. To our understanding, this is more to be ascribed to an overestimation of the number of galaxies of any mass predicted by their model than to the kind of prescription used to convert the gas into molecular.

We have also compared our results with those derived by \citet{fle+20} using the COLD GASS survey \citep{san+11,san+17}. Strictly speaking this comparison is only indicative because, as said above, the selection of the COLD GASS sample is not unbiased and there might be caveats not easy to understand. Our estimation lies a factor of 5 above \citet{fle+20}'s at low masses (log(M(\HII))$<$8.2) and lower of the same factor at 9.5$<$log(M(\HII))$<$10.

Finally we provide an estimate of the molecular mass density in the local Universe by integrating the functions shown in Figure~\ref{fig:MF}. We find a value $(1.6\pm0.6) 10^{7}M_\odot $Mpc$^{-3}$ and $(1.5\pm0.5) 10^{7}M_\odot $Mpc$^{-3}$ for our two estimated MFs with a constant and a luminosity-dependent \XCO,~respectively.
The values derived from the HRS at z=0 for both \MHc ~and \MHv~ agree within the errorbars and different by a factor of two with that of \citet{fle+20}, despite the fact that the latter authors use a \XCO~factor of $\alpha_{\rm CO}$ = 4.36~M$_\odot$/(K km s$^{-1}$ pc$^2$, a value 1.2 larger than that used in this paper ($\alpha_{\rm CO}$ = 3.6) \citep{bos+14b} 

\begin{figure}
  \centering{}
{\includegraphics[width=1.1\linewidth]{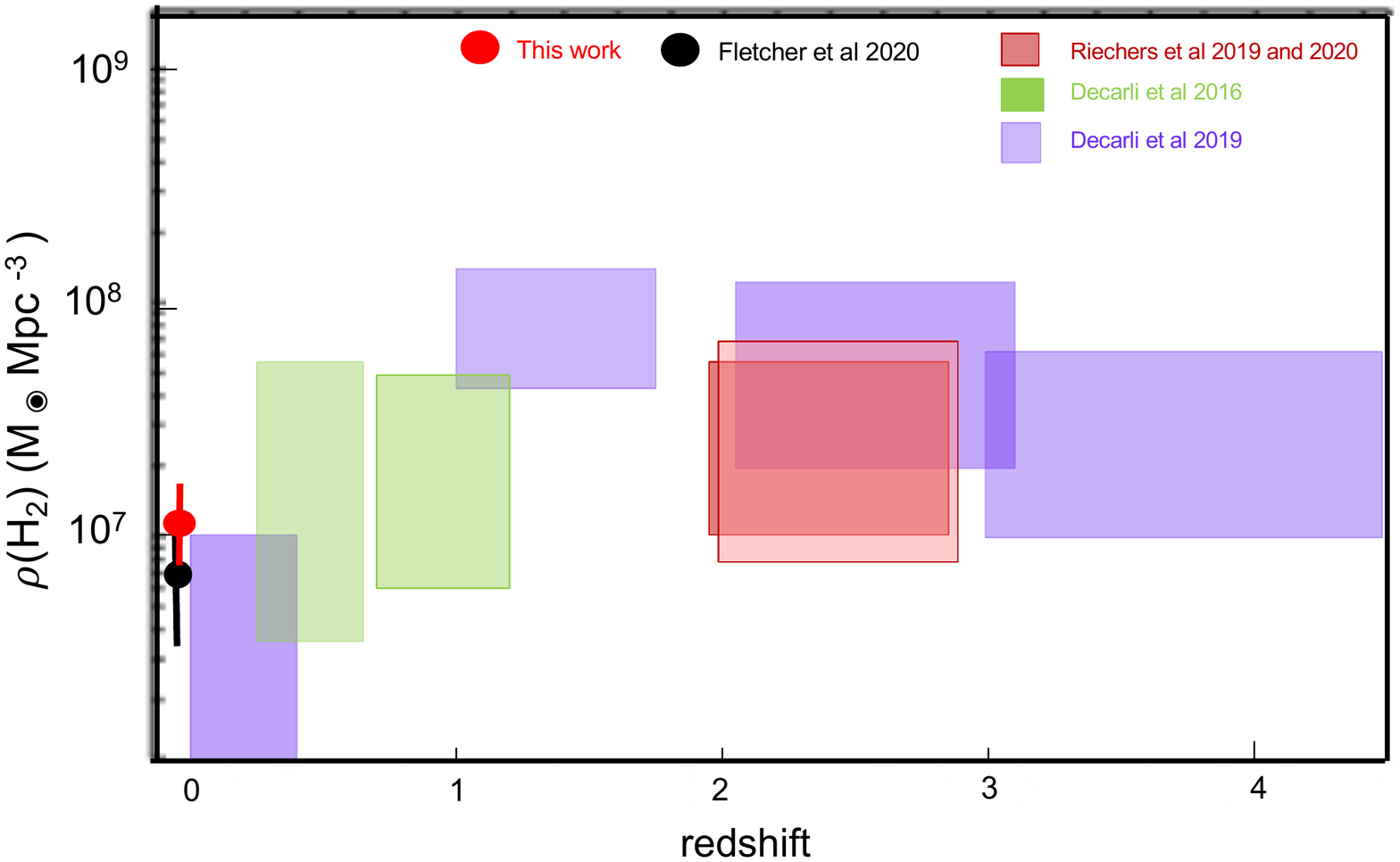}}
        \caption{The evolution of the molecular gas density as a function of redshift. Data are taken from \citet{dec+16,rie+19,dec+19,fle+20,rie+20}. The value from this work is shown in red.}
        \label{fig:Evolution}
\end{figure}

This result is placed in the Figure~\ref{fig:Evolution} where we have reported those inferred from higher redshift surveys carried out with a blind search in ALMA data cubes \citep{dec+16,rie+19,dec+19,rie+20}. 
Strictly speaking this graph allows only a qualitative comparison. The conversion to molecular gas mass from the CO line luminosity is highly dependent on various factors of which the dominant is the metallicity \citep{bol} but even in the Local Universe also the environment plays a role \citep[for instance, see][]{cic+18}. The status of the interstellar medium and therefore the properties of the CO-traced molecular gas at larger redshift is highly uncertain and the assumptions made to infer a molecular gas mass for the entire galaxy do not stand on solid ground \citep[see i.e.][]{pap+18}. Beside the \XCO~ conversion factor, $\alpha_{\rm CO}$ = 3.6~M$_\odot$/(K km s$^{-1}$ pc$^2$ used in both ALMA surveys \citep{dec+16,dec+19,rie+19,rie+20}, to infer the CO(1-0) line luminosity \citet{dec+19,rie+19} need to convert the
observed values CO(2-1), CO(3-2) CO(4-3) into the CO(1-0) making an assumption on the CO SLED. The used template is that of a standard CO SLED of star forming main sequence galaxies. Note that this latter argument is not any longer valid for the values reported in Figure~\ref{fig:Evolution} in the redshift range $z=2-3$ by \citet{rie+20} who have measured the CO(1-0) line emission with the VLA.
\par\noindent
An additional source of uncertainties of difficult quantification is the Cosmic variance associated to the small sky areas samples by the ALMA surveys.
\par\noindent
This means that the values of the molecular mass densities at various redshift is highly uncertain. Within the (large) errorbars the inferred values of the molecular mass density in the local Universe does not different from the values derived in surveys targeting higher redshift galaxies. There is no evidence at present of a large evolution of the molecular gas from $z$=0 up to $z\sim$4.



\begin{acknowledgements}
This research has made use of data from the HRS project. HRS is a Herschel Key Programme utilising guaranteed time from the 
SPIRE instrument team, ESAC scientists and a mission scientist. The HRS data was accessed
through the Herschel Database in Marseille (HeDaM - http://hedam.lam.fr) 
operated by CeSAM and hosted by the Laboratoire d’Astrophysique de Marseille.

P.A. warmly thanks the NRO staff at the Nobeyama Observatory for their fundamental help in performing the observations. She warmly thanks NAOJ-Chile for her stay at Mitaka where this work has been started.

This research benefited from discussions happened at the MIAPP workshop “Galaxy Evolution in a New Era of HI Surveys” held in 2019 and supported by the Munich Institute for Astro- and Particle Physics (MIAPP) which is funded by the Deutsche Forschungsgemeinschaft (DFG, German Research Foundation) under Germany Excellence Strategy - EXC-2094 -- 390783311.

This work makes use of SW routines contained in the $R$-package, $R$ is available as Free Software under the terms of the Free Software Foundation’s GNU General Public License in source code form, see 
${\rm https://www.r-project.org/about.html}$ 

This publication makes use of data products from the Two Micron All Sky Survey, which is a joint project of the University of Massachusetts and the Infrared Processing and Analysis Center/California Institute of Technology, funded by the National Aeronautics and Space Administration and the National Science Foundation.

\end{acknowledgements}

\end{document}